\documentclass{elsartL}
\usepackage{graphicx}
\usepackage{amssymb}
\usepackage{amsmath}
\usepackage{physics}
\usepackage{mathrsfs}
\usepackage{mathtools}
\usepackage{appendix}
\newcommand{\avg}[1]{\left< #1 \right>} 

\begin{document}

\begin{frontmatter}
\title{COVID-19: On the quarantine duration after short visits to high-risk regions}
\author{Evangelos Matsinos}
\begin{abstract}A simple Monte-Carlo method will be put forward herein, to enable the extraction of an estimate for the quarantine duration, applicable to visitors to high-risk regions. Results will be obtained on the basis 
of an analysis of the upper tail of the cumulative distribution function of the time span between the departure of the travellers from the place where the infection occurs and the time instant when COVID-19 infections may 
currently be detected. As expected, the quarantine duration is a decreasing function of the fraction of the infected travellers, which one is prepared to identify as `acceptable risk'. The analysis suggests that a maximal 
$5~\%$ risk (of new infections originating from subjects who become infective after their quarantine is lifted) may be associated with a minimal quarantine duration of about eight days, $1~\%$ with about twelve, and $0.1~\%$ 
with about sixteen. Unless the distribution of the duration of short (typically, up to three weeks) travels departs significantly from the shape assumed in this study, the results of the present analysis do not provide support 
for the plans to shorten the quarantine duration of about ten days to two weeks, which currently applies to travellers entering most European countries from regions with a high risk of infection.\\
%
\end{abstract}
\end{frontmatter}

\section{\label{sec:Introduction}Introduction}

The advent of 2020 brought humanity to the era of the Coronavirus (COVID-19) pandemia. This infectious disease, caused by the `Severe Acute Respiratory Syndrome Coronavirus 2' pathogen (SARS-CoV-2), poses a global threat to the 
provision of basic medical care, as well as to economic growth. As Gita Gopinath, the Economic Counsellor and Director of the Research Department at the International Monetary Fund, wrote on April 14 in an article entitled `The 
Great Lockdown: Worst economic downturn since the Great Depression': ``This is a truly global crisis as no country is spared. Countries reliant on tourism, travel, hospitality, and entertainment for their growth are experiencing 
particularly large disruptions. Emerging market and developing economies face additional challenges with unprecedented reversals in capital flows as global risk appetite wanes, and currency pressures, while coping with weaker 
health systems, and more limited fiscal space to provide support. Moreover, several economies entered this crisis in a vulnerable state with sluggish growth and high debt levels.'' \cite{Gopinath2020}

To restrain the rapid dissemination of the disease, lockdowns were imposed in most European countries and in the United States between March and early June, as well as in several Asian countries somewhat earlier. By the end of 
September, with over one million deceased worldwide (and counting), the use of masks in enclosed spaces became mandatory, whereas the authorities miss no opportunity to remind us of social distancing and attentiveness to personal 
hygiene, in particular hand disinfection. The effort towards carrying out tests on the verge of (what currently appears to be) the second wave of the pandemia intensified, isolation and quarantine entered our daily vocabulary, 
and contact-tracing applications were put in place to provide assistance in combatting the ramifications of new infections.

As the economic aspects of the mitigation measures make front-page news, another important issue is frequently underrated. Isolation, quarantine, and distance-learning/remote-working practices are expected to have a more lasting 
impact on personalities which are in the early development phases, i.e., on children and on adolescents. The elementary force in our society is the interaction between the individuals. This interaction shapes our social conduct 
and provides us with the necessary experience to cope with future situations arising in our dealings with other individuals or groups. Isolation, quarantine, and distance learning for the children and the adolescents imply that 
their only nonvirtual interactions are those with their immediate environment, presumably with the family, which comprise their `already known'. The likelihood that this `confinement within the known and alienation from the unknown' 
might leave an indelible mark upon the social competence of the minors should not be underestimated.

This technical note relates to the duration of the quarantine for visitors to high-risk regions. At this time, a quarantine of fourteen days applies to many European travellers when they return home from such regions. In the 
early phases of what appears to be the beginning of another critical period in relation to this disease, several European countries, i.e., the Netherlands, Norway, and Switzerland, have in place (since a few months) a ten-day 
quarantine, whereas other countries, e.g., Spain and Poland, plan on shifting to a quarantine of the same duration. With over $30\,000$ deceased and rising numbers of new infections for the last two months, France appears to 
vacillate between policies, contemplating a seven-day quarantine. Germany has announced plans to shorten the quarantine duration substantially, to five days, which according to virologist Christian Drosten would be ``sufficient''. 
There are also those who express disapproval of such plans, believing that ``halving the quarantine period would be negligent.'' \cite{Gallati2020}

Quarantine currently applies to
\begin{itemize}
\item[a)] visitors to high-risk regions, who subsequently travel to another destination,
\item[b)] subjects who have had a contact with a diagnosed or suspected case, and
\item[c)] subjects who have been diagnosed positive for COVID-19.
\end{itemize}
As aforementioned, this work pertains to case (a) above.

The goal herein is to investigate whether a simple procedure for determining the quarantine duration could be established from known facts, as well as from reasonable assumptions about the distribution of the duration of short 
travels, i.e., of travels lasting up to three weeks. A straightforward Monte-Carlo (MC) approach will be put forward as the most efficient manner to study such effects.

In this work, a test will be assumed to be reliable if its outcome is reproducible and accurate. The effectiveness of a COVID-19 test will be associated with the minimal viral load which is required for the confirmation of an 
infection. In this respect, tests will become more effective in the future, as (compared to the present time) they are expected to become capable of confirming an infection on the basis of smaller viral loads. From now on, the 
use of the term `an effective COVID-19 test' will refer to the most effective test available at the time for the detection of a COVID-19 infection, which - in addition - yields a reliable result within a few hours.

\section{\label{sec:Method}Method}

The travel histories will be generated as follows.
\begin{itemize}
\item The duration of travel will be sampled from a rescaled beta distribution between one and twenty-one days. The parameters of the beta distribution will be selected in such a way as to place the average of the distribution 
at about twelve days, which was the average duration of trips for German travellers in 2019 \cite{Statista}.
\item Each of the travellers will be infected at a random time instant during the travel: three models will be followed. The probability of infection will be assumed to be
\begin{enumerate}
\item constant within the entire duration of each travel,
\item increasing with time (i.e., from arrival to departure) according to a linear model, and
\item increasing with time (i.e., from arrival to departure) according to a logarithmic model.
\end{enumerate}
The uniform model represents travellers who are equally attentive to the observation of the precautions or of the mitigation measures throughout their travel, whereas the linear and logarithmic models take account of the 
psychological effect of habituation, i.e., of the reduction in one's response to a recurrent stimulus: it may be argued that, on average, the visitors tend to be more attentive at the early stages of the travel, relaxing their 
attention as time goes by (and `nothing happens').
\item The incubation interval for each subject will be sampled from a Weibull distribution.
\item The time instant at which each infected subject would test positive will be randomly selected within the fraction of the incubation interval in which the subject is believed to be infective (presymptomatic transmission).
\end{itemize}

\begin{figure}
\begin{center}
\includegraphics [width=12.5cm] {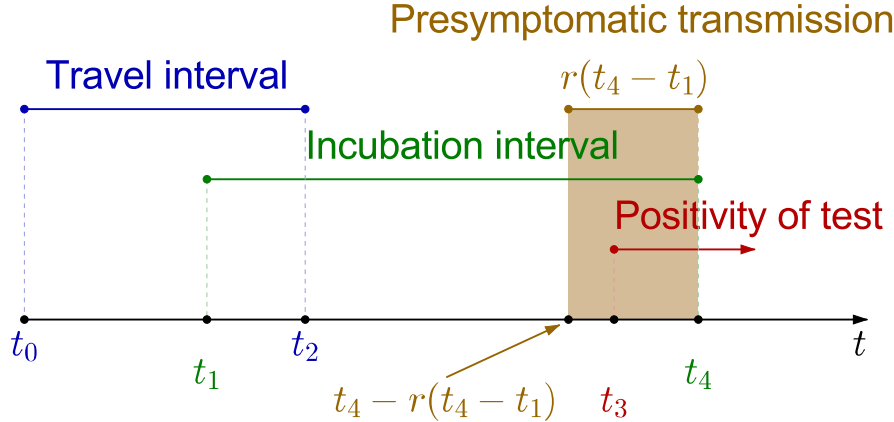}
\caption{\label{fig:Times}The milestones in each travel history: the time instant $t_0$ corresponds to the arrival of the traveller at the location of infection; $t_1$ represents the time instant of infection; $t_2$ marks the 
departure of the traveller from the location of infection, taken (for the sake of simplicity) to be also the time instant of the arrival of the traveller at the final destination; $t_3$ represents the time instant when the 
traveller, if tested via an effective COVID-19 test, would be found infected; finally, $t_4$ represents the onset of symptoms. The interval between the time instants $t_4 - r (t_4 - t_1)$ and $t_4$, shown highlighted in the 
figure, represents the time span within which the infected subject may transmit the disease, though no symptoms have been developed yet (presymptomatic transmission); $t_3$ is randomly selected within the highlighted interval.}
\vspace{0.35cm}
\end{center}
\end{figure}

The important time instants in each travel history are presented in diagrammatic form in Fig.~\ref{fig:Times}. One million travel histories, ramdomised by seeding the random-number generator with the time of the start of each 
run, will be generated per case, i.e.,
\begin{itemize}
\item for each method used in the sampling of the time instant of infection (Section \ref{sec:Infection}) and
\item for each of the three representative values of the only parameter of this work (Section \ref{sec:Positivity}).
\end{itemize}
In this study, the unit of time is one day (d).

\subsection{\label{sec:Duration}On sampling the duration of travel}

My intention had first been to assume the broadest possible set of travellers, taking no account of the gender, age, ethnicity, etc., of the type of travel (i.e., professional or recreational), and of the mode of transport. 
The ideal procedure would be to acquire such data from several countries. However, the process of obtaining the data would have been lengthy, as several authorities would have to be contacted, would have to concur, and would 
have to communicate their data in simple forms.

To provide a faster solution, the duration of travel was assumed to follow a beta distribution, rescaled between one and twenty-one days. The beta distribution is defined in the domain $[0,1]$: $0$ was mapped to one day and $1$ 
was mapped to three weeks. The parameter $\beta$ of the beta distribution was set (arbitrarily) to $2$, whereas the parameter $\alpha \approx 2.65$ was chosen so that the average of the resulting distribution would emerge at 
$12.4$ days, which, according to Ref.~\cite{Statista}, was the average duration of travels for the German population in 2019.

Evidently, the MC generation also takes account of the short travels, which are predominantly of a professional, rather than of a recreational, nature. Being shorter, professional travels tend to generate infections closer to 
the departure from the location of infection. Therefore, (in comparison with recreational travellers) the professional travellers are expected (on average) to be found infected at a later time after their arrival at the final 
destination.

\subsection{\label{sec:Infection}On sampling the time instant of infection}

I first thought that the sampling of the time instant of infection from a uniform distribution between the time instants $t_0$ and $t_2$ of Fig.~\ref{fig:Times} would suffice. However, my intuition suggested that the effects 
of habituation should not be ignored: one could argue that, facing an unknown setting, the travellers tend to be cautious and attentive. As time goes by, experience is gained and growing familiarity with the new setting may 
lead to the relaxation of the attentiveness. Consequently, I decided to take account of such effects by introducing two additional sampling methods: in the first, the time instant $t_1$ will be sampled according to a linear 
model between $t_0$ and $t_2$ (probability of infection linearly increasing with time from $t_0$ to $t_2$); in the second, $t_1$ will be sampled from a logarithmic distribution.

The probability density function (PDF) of the linear model is given by
\begin{equation} \label{eq:EQ001}
f(x) = 2 \frac{a x + b}{a + 2 b} \, \, \, ,
\end{equation}
for $x \in [0,1]$; $x=0$ will be mapped to $t_0$, $x=1$ to $t_2$. Without loss of generality, it may be assumed that the parameters $a$ and $b$ satisfy: $b > 0$ and $a > -b$; in addition, either $a$ or $b$ (but not both) may be 
equal to $0$. (The linear model reduces to the uniform distribution for $a=0$.) The mean of the distribution is given by
\begin{equation} \label{eq:EQ002}
\avg{x} = \frac{2 a + 3 b}{3 (a + 2 b)}
\end{equation}
and its variance by
\begin{equation} \label{eq:EQ002}
\sigma^2 = \frac{a^2 + 6 a b + 6 b^2}{18 (a + 2 b)^2} \, \, \, .
\end{equation}
The linear modelling admits one parameter, taken to be $f(0)$. The quantity $2/f(0)-1$ for $f(0) \neq 0$ is the ratio of the probabilities of infection at $x=1$ and $x=0$. In the MC simulation, this parameter was set equal to 
$1/2$, implying that the infection at the end of the travel is three times more probable than it was at the beginning.

The PDF of the logarithmic model is given by
\begin{equation} \label{eq:EQ003}
f(x) = c a \ln (a x + b) \, \, \, ,
\end{equation}
where 
\begin{equation} \label{eq:EQ004}
c = \left( \ln \left( \frac{(a+b)^{a+b}}{b^b} \right) - a \right)^{-1} \, \, \, .
\end{equation}
The permissible parameter space is somewhat more complex than in case of the linear model. The positivity of the various arguments of the logarithmic function, as well as that of $f(x)$ for $x \in [0,1]$, enforce the conditions: 
$-b < a \leq 1-b$ for $0 < b \leq 1$ and $a \geq 1-b$ for $b > 1$. The value $x=0$ will be mapped to $t_0$, $x=1$ to $t_2$. The mean of the distribution is given by
\begin{equation} \label{eq:EQ005}
\avg{x} = \frac{c}{a} \Bigg( \frac{1}{2} \ln \left( \frac{(a+b)^{(a+b)^2}}{b^{b^2}} \right) - b \ln \left( \frac{(a+b)^{a+b}}{b^b} \right) - \frac{a}{4} \left( a - 2 b \right) \Bigg)
\end{equation}
and its variance by
\begin{align} \label{eq:EQ006}
\sigma^2 &= \frac{c}{a^2} \Bigg( \frac{1}{3} \ln ( \frac{(a+b)^{(a+b)^3}}{b^{b^3}} ) - b \ln ( \frac{(a+b)^{(a+b)^2}}{b^{b^2}} ) \nonumber \\
&+ b^2 \ln ( \frac{(a+b)^{a+b}}{b^b} ) + \frac{a}{3} \big( -\frac{a^2}{3} + \frac{ab}{2} - b^2 \big) \Bigg) - \avg{x}^2 \, \, \, .
\end{align}
The logarithmic modelling admits two parameters, taken to be the ratios $f(1/2)/f(0)$ and $f(1)/f(0)$. In the MC simulation, these parameters were set equal to $2.5$ and $3$, respectively. This implies that the infection at the 
end of the travel is three times more probable than it was at the beginning; and that the infection halfway through the travel is $2.5$ times more probable than it was at the beginning. Therefore, in comparison with the linear 
model, the probability of infection in the logarithmic model rises more steeply in the first half of the travel interval (and more moderately in the second).

\subsection{\label{sec:IncubationInterval}On sampling the incubation interval}

The determination of the distribution of the incubation interval is detailed in Appendix \ref{App:AppB}. The results of the fit of the Weibull distribution to the available data (see Table \ref{tab:Parameters}) will be used in 
the MC generation.

\subsection{\label{sec:Positivity}On sampling the time instant when COVID-19 infections may currently be detected}

Although not corroborated (not necessarily refuted as well) by Ref.~\cite{Walsh2020}, let me assume for the sake of the argument that the positivity of an effective COVID-19 test is associated with a viral load $\rho_{\rm pos}$, 
whereas the development of symptoms is associated with a viral load $\rho_{\rm ons}>\rho_{\rm pos}$. Assuming an exponential growth, the viral load at time $t$ after the infection satisfies: $\rho (t) = \rho_0 2^{t/\tau}$, where 
$\tau$ (very likely, a subject-dependent quantity) is obviously the doubling time of the virus. Evidently, the earliest time instant $t_{\rm pos}$ for the positivity of the test is: $t_{\rm pos} = \tau \ln (\rho_{\rm pos}/\rho_0)/\ln (2)$. 
Similarly, the onset of symptoms occurs at $t_{\rm ons} = \tau \ln (\rho_{\rm ons}/\rho_0)/\ln (2)$. Therefore, $t_{\rm pos}/t_{\rm ons} = \ln (\rho_{\rm pos}/\rho_0) / \ln (\rho_{\rm ons}/\rho_0)$. In this naive picture, the 
time span between the infection and the positivity of the test is a constant fraction of the incubation interval.

The positivity of an effective COVID-19 test will be associated in this study with the presymptomatic transmission, for which substantial evidence has emerged \cite{Tindale2020,Ganyani2020,Wei2020,Arons2020,Ren2020,Zhang2020a,Han2020,Buitrago2020}. 
The interval $r (t_4-t_1)$, corresponding to the presymptomatic transmission, is shown highlighted in Fig.~\ref{fig:Times}, and (according to the literature) it does not exceed a typical duration of two to four days. The time 
instant $t_3$ of Fig.~\ref{fig:Times} at which the subject is found infected will be randomly selected within the interval $r (t_4-t_1)$. A linear model will be assumed when sampling $t_3$, increasing with time: this implies 
that the times close to the start of the presymptomatic transmission (i.e., to the time instant $t_4-r(t_4-t_1)$ in Fig.~\ref{fig:Times}) will be given less weight (than subsequent time instances up to time $t_4$). This 
behaviour is in general concordance with the results of Ref.~\cite{Kucirka2020}, though it is assumed herein that the effective COVID-19 test confirms infection with $100~\%$ success rate on the day of the onset.

\subsection{\label{sec:Tests}On the reliability of the COVID-19 tests}

A brief review of the currently-available COVID-19 testing options may be found in Ref.~\cite{Chang2020}.
\begin{itemize}
\item The fastest and simplest test is the antigen-antibody/serological test. It requires a few droplets of blood or serum, and targets the antibodies (immunoglobulin M and G) created by the organism as a response to the COVID-19 
infection. It takes about $10$ min for the test to yield results, but its accuracy is low ($50-70~\%$), in particular during the early phases of the infection (not enough antibodies for detection).
\item The test utilising the Polymerase Chain Reaction (RT-PCR), a method established in the mid 1980s for the amplification of small amounts of DNA to the level which would make a study feasible, targets the genes of the 
proteins of the virus. The rRT-PCR (real-time Reverse Transcription Polymerase Chain Reaction) version of the test needs about $4-6$ h to yield results, and requires effort and facilities. Worse still, the interpretation of the 
test results is not clear-cut: the diagnosis rests upon a comparison of the outcome of each test with a threshold value (Ct value) which does not seem to be sufficiently well-known \cite{Chang2020}. Also relevant is the issue 
of what the diagnosis would be in borderline situations, on either side of the assumed Ct threshold. In addition, though the test is described in Ref.~\cite{Chang2020} as ``highly accurate'' ($97~\%$), recent studies 
\cite{Kucirka2020,Woloshin2020} report \emph{glaring} failures even on the date of the onset of symptoms!\\
To reduce the probability of false negatives when using the rRT-PCR method, another RT-PCR test (the Pancoronavirus RT-PCR) has been established, first aiming at the identification of any of the Coronaviruses, then (if the result 
of the first test is positive) specifically searching for COVID-19. The test takes up to $24$ h to yield results.
\item The last method identifies the virus following the lengthy process of a cell culture, and may take up to two weeks to yield results (an expedited version of the test is also available). The test can be carried out only 
in strictly controlled environments, it entails infection risks for the examiners, but it is considered to be the most reliable at this time.
\end{itemize}

From all the above, it may be deduced that a reliable, cheap, and fast test for COVID-19 is currently unavailable. To provide a remedy for the large amount of false negatives (the highest risk in tests, as infective subjects 
are unaware of the risk they pose to others), some authors \cite{Ramdas2020} promulgate the repetitive testing of subjects: without doubt, this might be the way to go in case of a cheap and fast test (e.g., of the antigen-antibody/serological 
test), available to all those who need (or even wish) to be tested. Given the limited availability of the COVID-19 tests (in comparison with the population in a country, the number of test kits still remains small), the number 
of subjects who could be repeatedly tested will inevitably have to be cut down by the number of repetitions of the test per subject.

This section has been added for the sake of completeness. As the time instant when an infection is confirmed is sampled herein from the $[t_4 - r (t_4-t_1), t_4]$ domain, there is no room for the wrong identification of an 
infected subject as uninfected (false negative). In this respect, there can be no doubt that the quarantine durations extracted herein are on the optimistic side (i.e., smaller than they would have been, had a more realistic 
approach been assumed in relation to the effectiveness of the average COVID-19 test carried out in the field). In practice, this realisation reinforces further the conclusions of this work (see subsequent sections) that the 
shortening of the quarantine duration is not advisable at this time.

\section{\label{sec:Results}Results}

Nine runs were made, each generating one million travel histories using as input:
\begin{itemize}
\item the duration of travel, as described in Section \ref{sec:Duration};
\item the time instant of infection sampled from a uniform, a linear, and a logarithmic distribution, as described in Section \ref{sec:Infection};
\item the incubation interval, as described in Section \ref{sec:IncubationInterval}; and
\item the time instant when an effective COVID-19 test detects infection, as described in Section \ref{sec:Positivity}.
\end{itemize}
In the last case, the parameter $r$ of Section \ref{sec:Positivity} (fraction of the incubation interval within which the infected subject is infective) was set equal to $0.15$, $0.20$, and $0.25$. Therefore, three choices of 
the distribution from which the time instant $t_1$ is sampled $\times$ three choices of the $r$ value make nine runs in total.

\begin{figure}
\begin{center}
\includegraphics [width=12.5cm] {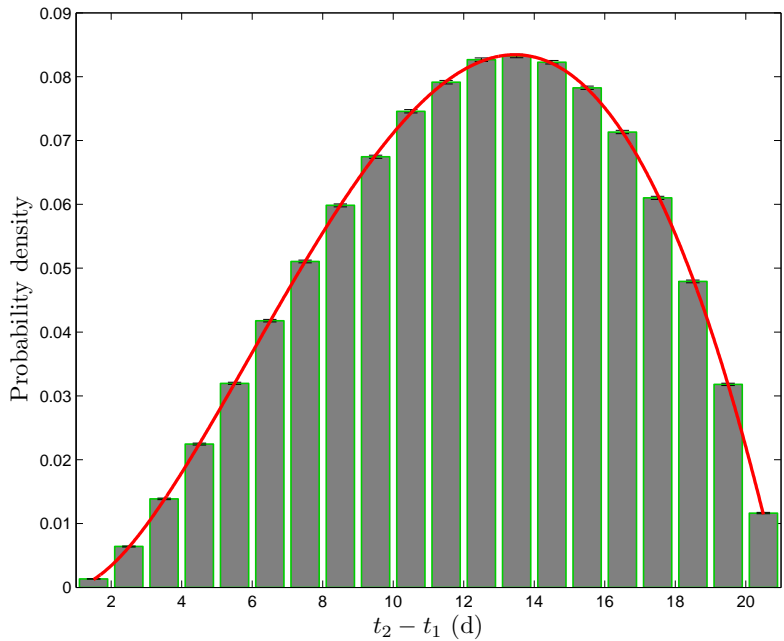}
\caption{\label{fig:TravelDurationDistribution}The distribution of the duration of travel. The histogram has been obtained from one million generated events. The red curve is the underlying beta distribution, rescaled to the 
domain between one and twenty-one days (see Section \ref{sec:Duration}).}
\vspace{0.35cm}
\end{center}
\end{figure}

The distribution of the duration of travel is shown in Fig.~\ref{fig:TravelDurationDistribution}. The incubation interval is sampled from the Weibull distribution displayed in Fig.~\ref{fig:IncubationIntervalDistribution}.

\begin{figure}
\begin{center}
\includegraphics [width=12.5cm] {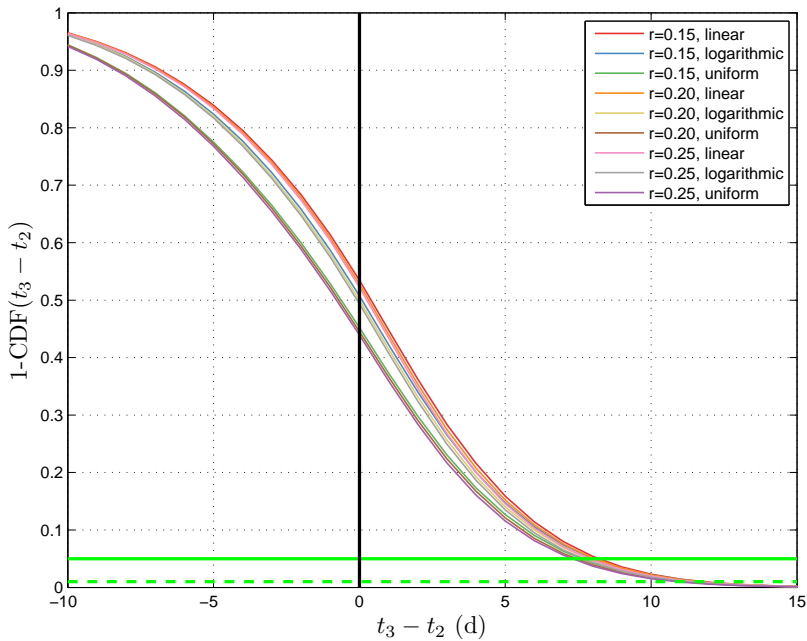}
\caption{\label{fig:Fraction}The upper tail of the cumulative distribution function of the time span between the arrival at the final destination and the time instant when an effective COVID-19 test detects infection, i.e., the 
quantity $t_3-t_2$ of Fig.~\ref{fig:Times}. The green horizontal solid and dashed lines mark the $5$ and $1~\%$ levels, respectively.}
\vspace{0.35cm}
\end{center}
\end{figure}

\begin{figure}
\begin{center}
\includegraphics [width=12.5cm] {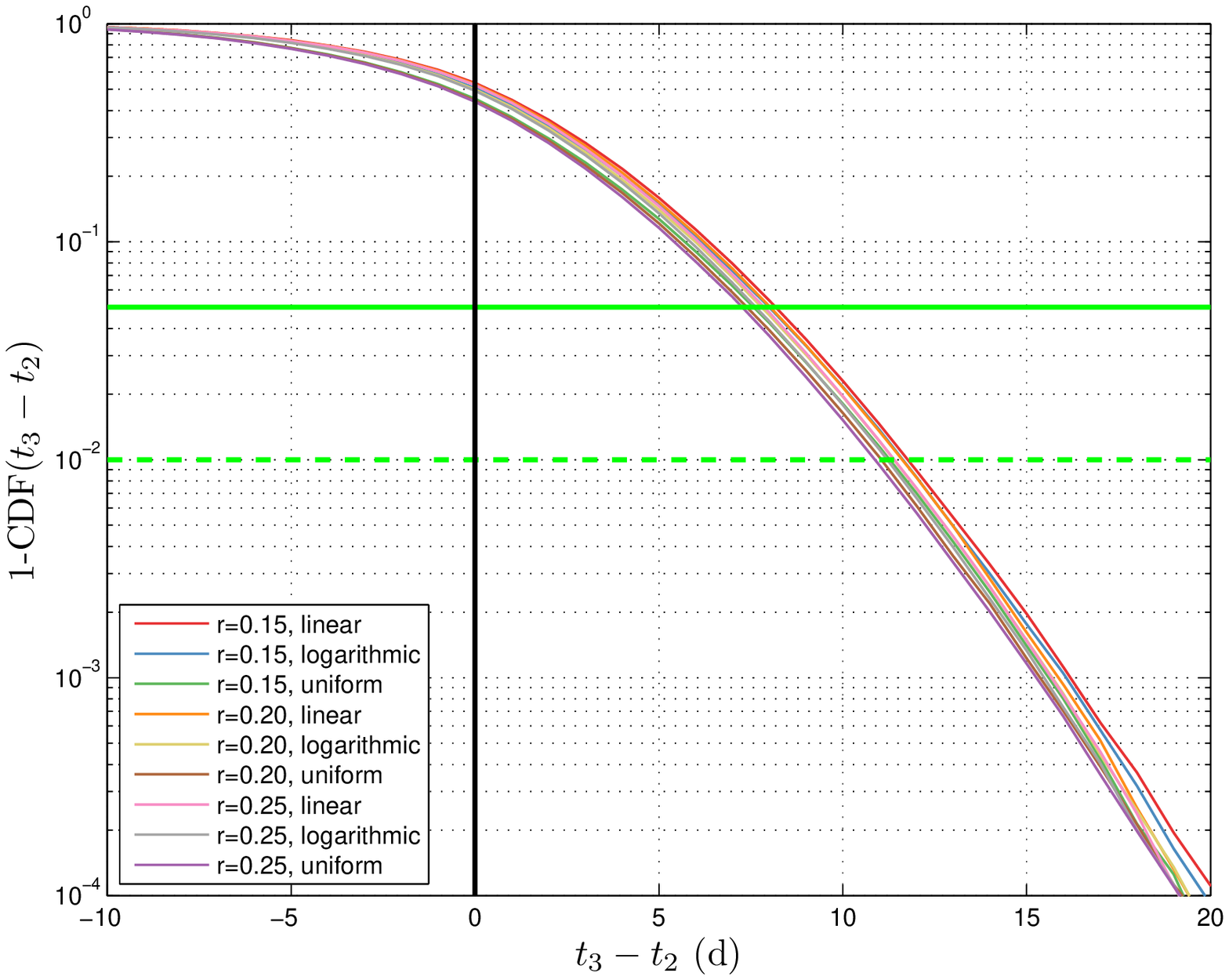}
\caption{\label{fig:FractionLogarithmic}Same as Fig.~\ref{fig:Fraction} for a logarithmic vertical scale.}
\vspace{0.35cm}
\end{center}
\end{figure}

The time span between the arrival at the final destination and the time instant when an effective COVID-19 test detects infection, i.e., $t_3-t_2$ in Fig.~\ref{fig:Times}, was histogrammed. The simulation leaves no doubt that 
about half of the travellers should be found infected before arrival at the final destination; therefore, they should not be allowed to travel (in principle). Those who make it to the final destination test negative (upon 
arrival) because (in the naive picture of Section \ref{sec:Positivity}) the viral load has not yet reached the detection level $\rho_{\rm pos}$. The upper tail of the cumulative distribution function of the time span $t_3-t_2$ 
is given in Figs.~\ref{fig:Fraction} (linear scale) and \ref{fig:FractionLogarithmic} (logarithmic scale, to allow for the details in the right tail of the distribution to appear more clearly).

\section{\label{sec:Conclusions}Discussion and conclusions}

This work relates to the mitigation measure of the quarantine, imposed on visitors to places where the probability of infection with the COVID-19 virus is sizeable; the authorities in each country issue their own lists of such 
high-risk regions. A Monte-Carlo (MC) method has been put forward, to enable the extraction of an estimate for the minimal quarantine duration from an analysis of the upper tail of the cumulative distribution function of the time 
span between the arrival of infected subjects at the final destination and the time when they would be found infected by an effective COVID-19 test.

A quarantine duration of eight days misses about $5~\%$ of the infected travellers. To be able to cut this fraction down to $1~\%$, a quarantine of twelve days appears to be unavoidable. A quarantine duration of slightly 
over two weeks would reduce this fraction further to about one per mille.

It must be borne in mind that the quarantine durations, given in this report, are minimal: it is unlikely that an ostensibly healthy person will subject him/herself to testing (at the right moment) before the onset of symptoms 
(unless, of course, that person suspects an infection due to contacts with subjects who have already been found infected). At this time, even in case that an infected subject does decide to take the test as a precautionary 
measure, the chances of a false-negative outcome do not appear to be neglectable \cite{Kucirka2020,Woloshin2020}.

The time span during which an infected subject remains infective does not enter the present study; the quarantine durations, extracted in this work, are independent of this quantity. This is because relevant to this study are 
only the events prior to the onset of symptoms. After the appearance of symptoms, another quarantine needs to be imposed, for a time span which (from what I have seen in the literature) is still not sufficiently well-known.

Not considered herein are the effects induced by asymptomatic subjects. As they have little to no indication that they have been infected, it is unlikely that these subjects will decide to subject themselves to testing. 
Regarding the asymptomatic subjects, relatively little is currently known. For instance, it has not been established for how long such subjects remain infective, as well as whether the infection probability is the same when a 
susceptible subject comes into contact with a symptomatic or with an asymptomatic subject. Also poorly known is the fraction of all infected subjects they represent: although Ref.~\cite{Matsinos2020} extracted from the 
available data (see relevant references therein) a population of about $17~\%$ of all infected subjects, there have been reports favouring significantly higher fractions.

Also not considered in this study is the impact of the quarantine duration on the economy. Without doubt, the determination of the optimal quarantine duration involves a difficult trade-off between its impacts on the public 
health, as well as on the economic growth in each country: to find the balance point is the challenge which most countries face at this time. My sole objection to decisions about the quarantine duration is when they are 
presented to the public as the outcome of the assessment of the former risks (impact on health), while they have been made after also taking the latter (impact on the economy) into account.

Provided that the distribution of the duration of travel of this work does not significantly depart from the true one, an inescapable conclusion may be drawn from the results of the MC simulation: the shortening of the 
quarantine duration, currently contemplated by the authorities in several European countries, cannot be supported.

Regarding the incubation interval (see Appendix \ref{App:AppB}), I would be very much interested in enhancing the database of subjects. I would also welcome data on the distribution of the duration of travels from reliable 
sources. I am interested in mixed samples: no gender/age/nationality discrimination, including all types of travels (professional/recreational) and (ultimately) all means of transport.

\begin{ack}
I am indebted to Karin M{\"o}lling for giving me the idea of this work, as well as for several discussions on issues of the dissemination of the COVID-19 disease.

I acknowledge numerous stimulating discussions with Valentina Viego, who was also more than willing to communicate to me the incubation-interval data used in Ref.~\cite{Viego2020}. In addition, I would like to thank Valentina 
for the ideas she gave me on the false-negative results of the COVID-19 tests: they might be pursued in a future work.

Finally, I also thank Natalie Linton, Marco Ajelli, Jantien Backer, and Jun Zhao for answering my questions.

Figure \ref{fig:Times} has been created with CaRMetal \cite{CaRMetal}. The remaining figures have been created with MATLAB$^{\textregistered}$ (The MathWorks, Inc., Natick, Massachusetts, United States).

I have no affiliations with or involvement in any organisation, institution, company, or firm with/without financial interest in the subject matter of this work.
\end{ack}

\clearpage
\newpage
\appendix
\section{\label{App:AppA}On fitting probability density functions with infinite support to histogram data}

Let $f(x)$ be a probability density function (PDF) of the real variable $x \in \mathbb{R}_{>0}$. Examples of such distributions are the three which are routinely employed in Epidemiology: gamma, log-normal, and Weibull. The 
purpose of this appendix is to outline a meaningful procedure for extracting estimates for the parameters of such distributions from histogram data extending from $x=0$ to $x_u$, arranged in $n$ histogram bins. The content of 
the histogram bin $i$ for $1 \leq i \leq n$ will be denoted by $N_i$.

Frequently in studies, the various PDF forms are directly fitted to the probabilities $\tilde{p}_i \coloneqq N_i/N$, where $N=\sum^n_{i=1} N_i$. This approach is problematic for three reasons:
\begin{itemize}
\item[a)] the departure of $f(x)$ from linearity within each histogram bin is neglected;
\item[b)] there is an evident dependence of the results on the bin size of the histogram; and
\item[c)] the approach takes no account of the fact that the PDF has infinite support (whereas $x_u$, being the largest right-hand endpoint, obviously remains finite).
\end{itemize}

To remedy the first two drawbacks, the fitted probability $p^f_i$ should be obtained as the difference of the corresponding cumulative distribution function (CDF) at the two endpoints $a_i$ and $b_i$ of the histogram bin $i$:
\begin{equation} \label{eq:EQA001}
p^f_i = \int_{a_i}^{b_i} f(x) dx \, \, \, .
\end{equation}

To remedy the third drawback, a correction needs to be applied to the histogram data, taking account of the fact that the integral
\begin{equation} \label{eq:EQA002}
\alpha \coloneqq \int_{x_u}^{\infty} f(x) dx > 0 \, \, \, .
\end{equation}
In other words, the true probabilities $p_i$ differ from $\tilde{p}_i$, in that they should involve not only the sum of the observations $N$, but also the contributions to the distribution from the infinite interval $x>x_u$, 
i.e., from existing observations which have been omitted (e.g., because they are sparse) or which did not find their way to the database (e.g., due to limited statistics); whether the data has been right-censored or 
right-truncated is of no relevance. In practice, $N$ must be replaced by $N / (1 - \alpha)$ when assigning a probability to the histogram bin $i$:
\begin{equation} \label{eq:EQA003}
\tilde{p}_i \coloneqq \frac{N_i}{N} \to p_i \coloneqq \frac{N_i}{N_\infty} = \frac{(1 - \alpha) N_i}{N} = (1 - \alpha) \tilde{p}_i \, \, \, .
\end{equation}
The quantities $p_i$ are the true probabilities, to be compared in the optimisation with $p^f_i$ of Eq.~(\ref{eq:EQA001}). Following from the binomial distribution, the uncertainty $\delta p_i$ of the true probability $p_i$ is 
obtained via the formula:
\begin{equation} \label{eq:EQA004}
\delta p_i = \sqrt{\frac{p_i (1-p_i)}{N_\infty}} = (1 - \alpha) \sqrt{\frac{\tilde{p}_i \left( 1 - \tilde{p}_i (1 - \alpha) \right)}{N}} \, \, \, .
\end{equation}
Therefore, the true probabilities $p_i$ and their uncertainties $\delta p_i$ can be evaluated from the observations $N_i$ and from the upper tail $\alpha$ of the CDF.

When a $\chi^2$ minimisation function is used in the optimisation, the following formula applies:
\begin{equation} \label{eq:EQA005}
\chi^2 = N \sum_{i=1}^{n} \frac{\left( \tilde{p}_i - p^f_i / (1 - \alpha) \right)^2}{\tilde{p}_i \left( 1 - \tilde{p}_i (1 - \alpha) \right)} \, \, \, .
\end{equation}
For large values of $x_u$, $\alpha \to 0$, and Eq.~(\ref{eq:EQA005}) reduces to the better-known formula
\begin{equation} \label{eq:EQA006}
\chi^2 = N \sum_{i=1}^{n} \frac{\left( \tilde{p}_i - p^f_i \right)^2}{\tilde{p}_i \left( 1 - \tilde{p}_i \right)} \, \, \, .
\end{equation}
However, Eq.~(\ref{eq:EQA005}) makes no assumptions about the largeness of $x_u$ and should be used when fitting PDFs with infinite support to histogram data. The quantity $\alpha$ is obtained \emph{at each step of the 
optimisation} from the parameters of the corresponding distribution.

\newpage
\section{\label{App:AppB}The extraction of the distribution of the incubation interval from available data}

To the best of my knowledge, seventeen peer-reviewed studies have reported on the incubation interval. Eight of these works \cite{Backer2020,Linton2020,Lauer2020,Pung2020,Zhang2020,Sanche2020,Yang2020,Pak2020} have made the 
details of their data publicly available. The details of the data, used in one additional study \cite{Viego2020}, were also communicated to me by the first author. My efforts notwithstanding, it has not been possible to enhance 
the database further by also including the data of the remaining eight studies.

From the $996$ available subjects, selected were those with complete data regarding the exposure window (i.e., the start and the end dates of the exposure interval), as well as the date corresponding to the onset of symptoms. 
Only one correction was applied: if the date on which the onset of symptoms preceded the one corresponding to the end of the exposure window, then the latter was set equal to the former. The maximal incubation interval, admitted 
to the analysis, was $25$ days. (As there is no subject with a maximal incubation interval of $24$ days, that quantity was reset - automatically in the software application - to $24$ days.) A contribution from each subject to 
the distribution of the incubation interval, inversely proportional to the duration of the exposure window, was appended to the contents of the histogram bins contained within that window. Evidently, assumed was a uniform 
distribution of the probability of infection within each exposure window.

\begin{figure}
\begin{center}
\includegraphics [width=12.5cm] {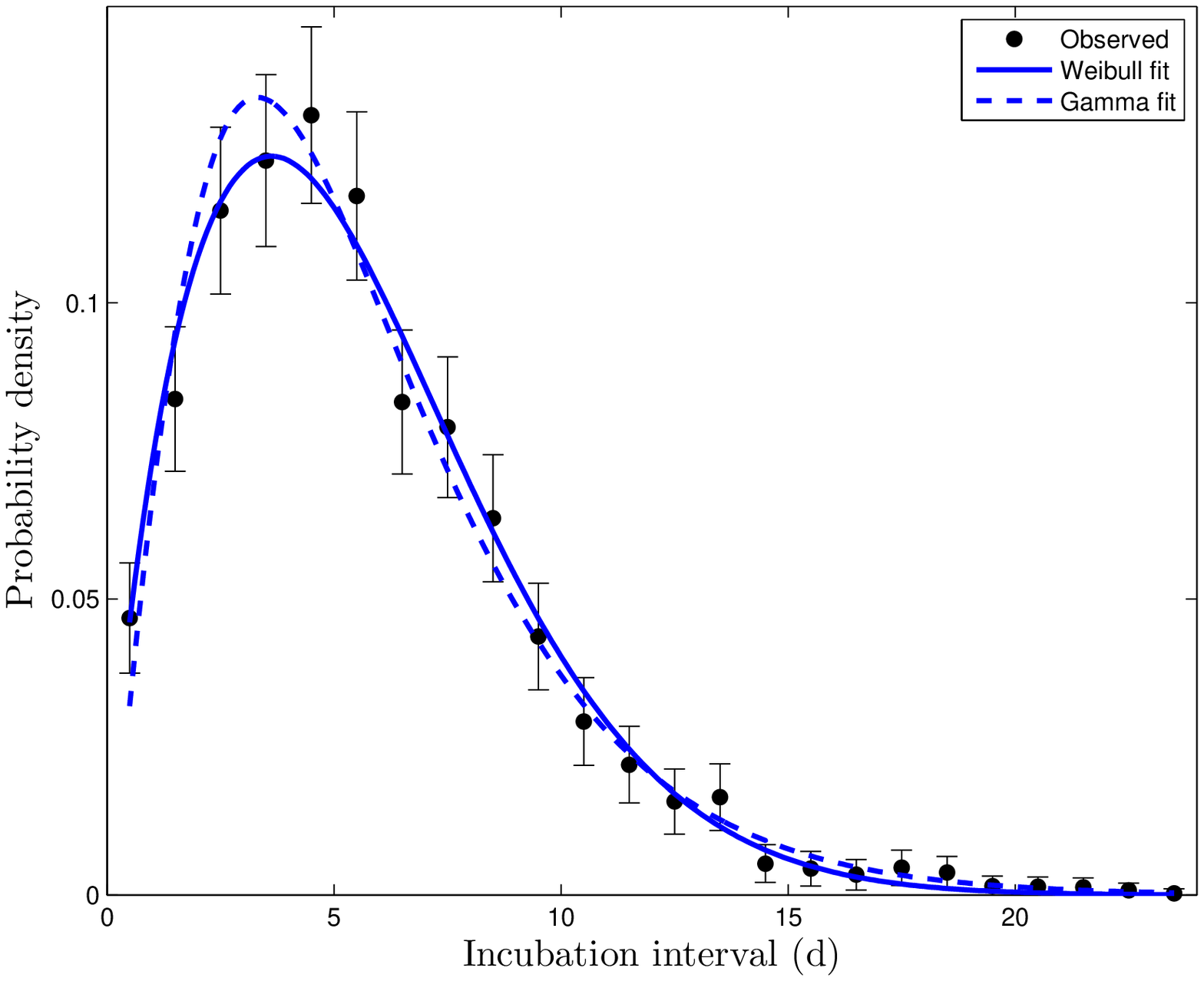}
\caption{\label{fig:IncubationIntervalDistribution}The distribution of the incubation interval, obtained from $516$ subjects with known details about the exposure window and the date of the onset of symptoms. The database was 
created from the publicly-available details given in Refs.~\cite{Backer2020,Linton2020,Lauer2020,Pung2020,Zhang2020,Sanche2020,Yang2020,Pak2020}, as well as from one set communicated to me by the first author of Ref.~\cite{Viego2020}.}
\vspace{0.35cm}
\end{center}
\end{figure}

After applying the corrections~\footnote{These corrections are negligible for the distribution dealt with in this section: in case of the Weibull fit, $\alpha \approx 6.43 \cdot 10^{-5}$; in case of the gamma fit, 
$\alpha \approx 8.69 \cdot 10^{-4}$.} outlined in Appendix \ref{App:AppA}, the data was submitted to a $\chi^2$-based optimisation, utilising three standard distributions: gamma, log-normal, and Weibull. As two free parameters 
enter these distributions, there are $22$ degrees of freedom (DoFs) in each fit. It was found that the results, obtained with the log-normal distribution, are not as good as those extracted with the other two distributions: with 
a p-value barely over $1~\%$ (namely, $1.17 \cdot 10^{-2}$), the log-normal distribution strives to accommodate the (relatively short) tail of the observed distribution of the incubation interval. On the other hand, the data 
is well described by the gamma and Weibull distributions. The fitted values and uncertainties of the parameters of these two distributions are detailed in Table \ref{tab:Parameters}. A plot of the data, along with the fitted 
values in case of the gamma and Weibull fits, is shown in Fig.~\ref{fig:IncubationIntervalDistribution}.

\begin{table}
{\bf \caption{\label{tab:Parameters}}}Fitted values and uncertainties of the parameters of the gamma and Weibull distributions, as they come out of the fit to the incubation-interval data from $516$ subjects with known details 
about the exposure window and the date of the onset of symptoms \cite{Viego2020,Backer2020,Linton2020,Lauer2020,Pung2020,Zhang2020,Sanche2020,Yang2020,Pak2020}. Shown are the results obtained from two (equivalent) modelling 
options in each case: the mean and rms of the two aforementioned distributions may be expressed in terms of the shape and scale parameters, and vice versa. The median in case of the fit using the Weibull distribution is about 
five days. The two scale parameters, as well as the means and the rms's of the distributions, are expressed in days.
\vspace{0.2cm}
\begin{center}
\begin{tabular}{|l|c|c|c|c|}
\hline
Distribution & Parameter & Fitted results & Parameter & Fitted results\\
\hline
\hline
\multicolumn{5}{|c|}{Standard parameterisation}\\
\hline
Gamma & Shape $a$ & $2.41^{+0.18}_{-0.16}$ & Scale $b$ (d) & $2.35^{+0.18}_{-0.17}$\\
Weibull & Shape $k$ & $1.675^{+0.066}_{-0.062}$ & Scale $\lambda$ (d) & $6.20 \pm 0.17$\\
\hline
\multicolumn{5}{|c|}{Parameterisation using the mean and the rms of the distribution}\\
\hline
Gamma & Mean (d) & $5.64 \pm 0.15$ & rms (d) & $3.64^{+0.16}_{-0.15}$\\
Weibull & Mean (d) & $5.54 \pm 0.15$ & rms (d) & $3.40 \pm 0.15$\\
\hline
\end{tabular}
\end{center}
\vspace{0.5cm}
\end{table}

There have been reports in the literature of longer (mean/median) incubation intervals than those appearing in Table \ref{tab:Parameters} of this work. Several of those values have been obtained after including in the analysis 
subjects with partially-known exposure windows, e.g., subjects with unknown start date of the exposure. To accommodate such subjects in the sample and gain in statistics, assumptions need to be made about their most probable 
time instant of infection. I suspect that the larger values of the incubation interval, obtained in those works, are due to assumptions about the distribution of the probability of infection in an infinite (in principle) 
exposure window. On the contrary, as only subjects with well-known details have contributed to the distribution shown in Fig.~\ref{fig:IncubationIntervalDistribution}, this work rests upon no such assumptions.

Regarding the comparison of results, obtained in several studies from different modelling options of the same data, I have one comment. Many authors favour modelling options on the basis of the application of the Akaike 
information criterion (AIC) \cite{Akaike1974}. My stand is that such complexity is entirely uncalled for in the problem dealt with in this section. In case of normally-distributed residuals, the maximisation of the likelihood 
is mathematically equivalent to the minimisation of the $\chi^2$ function. When the different modelling options make use of the same number of parameters to fit the same data, the resulting value of the AIC score is bound to 
be equal to the $\chi^2_{\rm min}$ value plus a constant! Several authors select the modelling option with the minimal AIC score, but refrain from addressing the significance of that choice. Well, yes, there will always be one 
modelling option which yields a better result. However, is that result significantly different to the second best?

There is only one established method (which I am aware of) for addressing the significance of the difference between two results obtained via a $\chi^2$-based method: it involves Fisher's ($F$) distribution. If the two results are: 
$\chi^2_1$ with $\nu_1$ DoFs and $\chi^2_2$ with $\nu_2$ DoFs, where $\chi^2_1 / \nu_1 > \chi^2_2 / \nu_2$, one first evaluates the ratio
\begin{equation} \label{eq:EQA007}
u = \frac{\chi^2_1 / \nu_1}{\chi^2_2 / \nu_2} \, \, \, .
\end{equation}
The quantity $u$ follows the $F$ distribution with $\nu_1$ and $\nu_2$ DoFs. In case of the incubation data of this section, the application of the AIC criterion would result in a difference of about $2.2$ between the gamma and 
the Weibull fits. On the basis of this score, many authors would probably hasten to consider the Weibull fit `superior'. However, the $u$ score of about $1.28$ for $22$ and $22$ DoFs yields the p-value of about $2.86 \cdot 10^{-1}$, 
hence no indication that the fit quality is significantly different in the two cases.

\end{document}